# Scalar Pair Production in the Aharonov-Bohm Potential

*G. Y. Shahin*[†] and *M. S. Shikakhwa*[‡]
Department of Physics, University of Jordan,
11942-Amman, Jordan

*Abstract*

In the framework of QED, scalar pair production by a single linearly polarized high-energy photon in the presence of an external Aharonov-Bohm potential is investigated. The exact scattering solutions of the Klein-Gordon equation in cylindrically symmetric field are constructed and used to write the first order transition amplitude. The matrix elements and the corresponding differential scattering cross-section are calculated. The pair production at both the nonrelativistic and the ultrarelativistic limits is discussed.

PACS numbers:

[†] ghassans@bethlehem.edu
[‡] moody@ju.edu.jo

## I. Introduction

Since the pioneering work of Aharonov and Bohm [1] about half a century ago, the systems in which charged particles interact with the vector potential of an infinitely long, thin magnetic string (*AB* potential) are still receiving considerable interest in the literature. In such systems, the non-local interaction of the charged particle with the magnetic field of the string leads, quantum mechanically, to observable physical effects despite the absence of Lorentz forces on the particle. The works [2, 3] provide an excellent review of the subject and its application in various areas.

Recently, works addressing issues other than the elastic scattering of charged particles off the AB potential had appeared. Serebryanyĭ *et al* [4] reported the differential cross section for Bremsstrahlung of non-relativistic particles in the *AB* potential in the dipole approximation. Gal'tsov *et al* [5] considered the synchrotron radiation by a relativistic scalar particle in the *AB* potential. Bremsstrahlung by spin-1/2 particles was considered in [6], and the cross section for the electron-positron pair production by a single photon in the *AB* potential was calculated in [7].

The present work was mainly motivated by the series of works [6-8]. In this work, we consider the production of a scalar particle-antiparticle pair in an *AB* potential by a single, linearly polarized photon. The exact differential cross section for the process is calculated and the limits of low and high-energy photons are discussed. Selection rules on the angular momentum of the resulting pair, that were reported for fermions [6,7], are also revealed in the scalar case. The framework of our calculations is the covariant perturbation theory. However, we expand the scalar field operators using, as basis, the exact scattering particle-antiparticle solutions of the Klein-Gordon (KG) equation in the *AB* potential, rather than using the free solutions of the free KG equation as basis as usually done (for the Coulomb potential, for instance) [9-11]. Such an approach is discussed in [12] and is used in [6,7]. This, in turn, makes the lowest non-vanishing contribution to the process of first order rather than second order. Calculations of the differential cross section of Bremsstrahlung and by a single photon and pair production for both scalar and spin one-half particles in the *AB* potential, using the free particle solutions as expansion basis, were also carried out [13, 14].

In Section II, we construct the particle and antiparticle solutions of the KG equation with coupling to the *AB* potential. The calculation of the differential cross-section for the scalar particle-antiparticle pair by a single, linearly polarized photon is carried out in Section III. Section IV discusses the limits for low and high-energy photons. Finally, we sum up and state our conclusions in Section V.

## II. The Exact Scattering Solutions to the Klein-Gordon Equation in the AB Potential

In the presence of an external *AB* vector potential field, $A_\mu$, one should make use of the minimal coupling in which the momentum operator $\hat{P}_\mu \to \hat{P}_\mu - \frac{e}{c} A_\mu$ (with $e = -|e|$). In this case, the interacting Klein-Gordon equation becomes:

$$(\hat{P}^\mu - \frac{e}{c} A^\mu)(\hat{P}_\mu - \frac{e}{c} A_\mu)\psi(\vec{r},t) = M^2 c^2 \psi(\vec{r},t) \qquad (1)$$

In this paper, we will consider only the idealized case (pure *AB* case) of an infinitesimally thin, infinitely long straight magnetic tube. In the absence of Coulomb potential and in cylindrical coordinates described by $(\rho, \varphi, z)$, it can be readily shown that the external vector potential has only an angular component given by

$$eA_\varphi = \frac{e\Phi_{en}}{2\pi\rho} = \frac{-\Phi_{en}}{\Phi_0} \frac{\hbar c}{\rho} = \frac{-\hbar c}{\rho} f \qquad (2)$$

where $\Phi_{en} \equiv f\Phi_0$ is the enclosed magnetic flux through the tube,

$\Phi_0 \equiv -2\pi\hbar c/e$ is the magnetic flux quantum,

$\rho = \sqrt{x^2 + y^2}$, and

$f = [f] + \delta$ with $[f]$ being the integer part of $f$, and $\delta$ is fractional quantity that produces all physical effects.

Substituting the above vector potential into Eq.(1), we have

$$\left[\frac{1}{c^2}\frac{\partial}{\partial t^2} - \frac{1}{\rho}\frac{\partial}{\partial \rho}\rho\frac{\partial}{\partial \rho} - \frac{1}{\rho^2}\frac{\partial^2}{\partial \varphi^2} - \frac{\partial^2}{\partial z^2} + \frac{M^2c^4}{\hbar^2} + \frac{f^2}{\rho^2} - \frac{2if}{\rho^2}\frac{\partial}{\partial \varphi}\right]\psi(\vec{r},t) = 0 \qquad (3)$$

We require $\psi(\vec{r},t)$ to be an eigenfunction of the following operators: the third component of linear momentum $\hat{P}_3$, the third component of the angular momentum $\hat{L}_3$, and the Hamiltonian $\hat{H}$:

$$\hat{P}_3\psi(\vec{r},t) = k_3\hbar\psi(\vec{r},t),\ \hat{L}_3\psi(\vec{r},t) = m\hbar\psi(\vec{r},t),\ \hat{H}\psi(\vec{r},t) = E_n\psi(\vec{r},t) \qquad (4)$$

where $\hbar k_3$, $\hbar m$ and $E_n = \hbar c\varepsilon_n$ are the eigenvalues of $\hat{P}_3$, $\hat{L}_3$ and $\hat{H}$ operators, respectively. In other words, $\{\hat{P}_3,\hat{L}_3,\hat{H}\}$ constitutes a complete set of commuting operators that are integrals of the motion. In this *ansatz*, the partial mode solution for $E_n > 0$, with normalization constant $N$, reads

$$\psi_{(+)}(\rho,\varphi,z;t) = Ne^{-i(c\varepsilon_n t - k_3 z - m\varphi)}R_m(\rho) \qquad (5)$$

where $R_m(\rho)$ obey the following radial KG equation:

$$\left[\frac{d^2}{d\rho^2} + \frac{1}{\rho}\frac{d}{d\rho} + k_\perp^2 - \frac{(f+m)^2}{\rho^2}\right]R_m(\rho) = 0. \qquad (6)$$

Eq.(6) is the usual form of Bessel equation of non-integer order $\tilde{m}_\pm \equiv m \pm f$. It has the general positive-energy solutions:

$$R_m(\rho) = F_m J_{|\tilde{m}_+|}(k_\perp\rho) + G_m J_{-|\tilde{m}_+|}(k_\perp\rho),$$

where $k_\perp$ is the component of $\vec{k}$ in the x-y plane.

Since $J_{\pm|\nu|}(x) \sim x^{\pm|\nu|}$ for $x \to 0$, $G_m$ must equal to zero if we insist on the regularity at the origin. Physically speaking, the irregular solutions must be eliminated since the scalar particle carries no magnetic moment. As a result, the particle suffers no interaction with the magnetic field at $\rho = 0$. In contrast, in the spinor case [15], there is an interaction between the spin and the magnetic moment. Therefore, the wave functions in the latter case do not vanish at $\rho = 0$. This leads to the problem of self-adjointness extension of the Hamilton operator [16-18].

The normalization constant $N$, can be determined by the following normalization condition [19] within a cubic box of volume $V = L^3$.

$$\frac{i\hbar}{2mc^2}\int[\psi^*_{(\pm)}(\vec{r},t)\frac{\partial \psi_{(\pm)}(\vec{r},t)}{\partial t} - \psi_{(\pm)}(\vec{r},t)\frac{\partial \psi^*_{(\pm)}(\vec{r},t)}{\partial t}]d^3\vec{r} = 1, \qquad (7)$$

However, the complete set of solutions of the interacting KG equation includes the negative-energy states $\psi_{(-)}$, in addition to the positive-energy states. These states, are - as usual - used to construct the antiparticle positive-energy states, described by $\psi_c$, through the charge-conjugation operation, in which $\psi_{(-)} \to \psi_c = \psi^*_{(-)}$ and replacing $e$ by $-e$.

Thus, the two normalized independent solutions are now the particle partial mode solutions, for $E_n = E^a > 0,$ given by

$$\psi_{(+)}^m(\rho,\varphi,z;t) = \sqrt{Mc^2/E_nV}\, e^{-i(c\varepsilon_n t - k_3 z - m\varphi)} J_{|\tilde{m}_+|}(k_\perp \rho), \qquad (8\text{-a})$$

and the antiparticle partial mode solutions, for $E'_{\bar{n}} = E^{\bar{a}} > 0,$ given by

$$\psi_c^{m'}(\rho,\varphi,z;t) = \sqrt{Mc^2/E'_{\bar{n}}V}\, e^{i(c|\varepsilon_{\bar{n}}|t - k'_3 z - m'\varphi)} J_{|\tilde{m}_-|}(k_\perp \rho). \qquad (8\text{-b})$$

The cylindrical partial mode solutions do not describe *incoming* and *outgoing* particles with definite linear momenta at infinity. In order to find out the scattering matrix-element, scattering solutions should be constructed. These solutions can be expressed as linear combinations of the partial modes of Eqs.(8-a,b) as

$$\Psi^a \equiv \Psi_{out} = \sum_{m=-\infty}^{\infty} c_m \psi_{(+)}^m(\rho,\varphi,z;t), \qquad (9\text{-a})$$

$$\Psi^{\bar{a}} \equiv \Psi_{in} = \sum_{m'=-\infty}^{\infty} c_{m'} \psi_c^{m'}(\rho,\varphi,z;t) \qquad (9\text{-b})$$

The coefficients $c_m$ are determined by using the fact that $\Psi_{in}$ must behave at large distance like a plane wave propagating in the direction $\vec{k}'$ plus an outgoing cylindrical wave. Likewise, $c_{m'}$ are determined since $\Psi_{out}$ must behave at large distance like a plane wave propagating in the direction $\vec{k}$ plus an incoming cylindrical wave [12]. It must be stressed that the incoming cylindrical wave of a particle, with angular momentum quantum number $m'$, can be viewed as an *outgoing* cylindrical wave of the corresponding antiparticle.

By making use of the familiar expansion of plane waves in terms of Bessel functions together with the asymptotic form of Bessel functions, it can be shown that the amplitudes $c_m$ and $c_{m'}$ are given, respectively, by:

$$c_m = e^{i\{m(\pi - \varphi_\perp) - \frac{\pi}{2}|\tilde{m}|\}} \text{ and } c_{m'} = e^{i\{-m'(\pi - \varphi'_\perp) + \frac{\pi}{2}|\tilde{m}'|\}} \qquad (10)$$

where $\varphi'_\perp$ and $\varphi_\perp$ are, respectively, the polar angles subtended by the outgoing momenta as $\rho \to \infty$. With this choice of the coefficients for the outgoing particle-antiparticle pair, the particle and aniparticle field operators can be expanded now in terms of the scattering states $\Psi^a$ and $\Psi^{\bar{a}}$ in a correct way [6-8], [12].

## III. The Matrix-Element Calculations

In order to calculate the matrix element for the production of a pair of massive particles by an incident high-energy massless particle such as the photon, it is necessary to find out, first fo all, the probability amplitude given by [19]

$$\vec{d} = \frac{1}{2Mc} \int d^3r\, e^{i\vec{\kappa}\cdot\vec{r}} \left( \left[\hat{\vec{P}}^* \Psi^{a*}(\vec{r})\right] \Psi^{\bar{a}}(\vec{r}) + \Psi^{a*}(\vec{r}) \left[\hat{\vec{P}} \Psi^{\bar{a}}(\vec{r})\right] \right) \qquad (11)$$

where
$\hat{\vec{P}} = -i\hbar\vec{\nabla} - (e/c)\vec{A}$ is the generalized momentum operator vector. $\Psi^a(\vec{r})$ and $\Psi^{\bar{a}}(\vec{r})$ are the time independent scattering wavefunctions of the outgoing particle and antiparticle, repectively, and given by

$$\Psi^a(\rho,\varphi,z) = \sqrt{Mc^2/E_n V}\, e^{ik_3 z} \sum_{m=-\infty}^{m=\infty} c_m e^{im\varphi} J_{|m+f|}(k_\perp \rho),$$

$$\Psi^{\bar{a}}(\rho,\varphi,z) = \sqrt{Mc^2/E'_{\bar{n}} V}\, e^{-ik'_3 z} \sum_{m'=-\infty}^{m'=\infty} c_{m'} e^{-im'\varphi} J_{|m'-f|}(k'_\perp \rho).$$

$\vec{\kappa}$ is the wavevector of the incoming photon, expressed in a coordinate system defined in Fig. 1, such that

$$\vec{\kappa}\cdot\vec{r} = \kappa\rho\sin\vartheta_\kappa \cos(\varphi-\varphi_\kappa) + \kappa z\cos\vartheta_\kappa.$$

The integrand can be viewed as a coupling of the charged current of the scalar particles to the external radiation field.

In this paper, a special coordinate system, described in Fig.1, will be followed. In the same figure, two polariztion vectors, $\hat{e}^{(\sigma)}$ and $\hat{e}^{(\pi)}$, are also shown that will be used later on.

After extensive manipulations, one gets the probability amplitude vector in the form

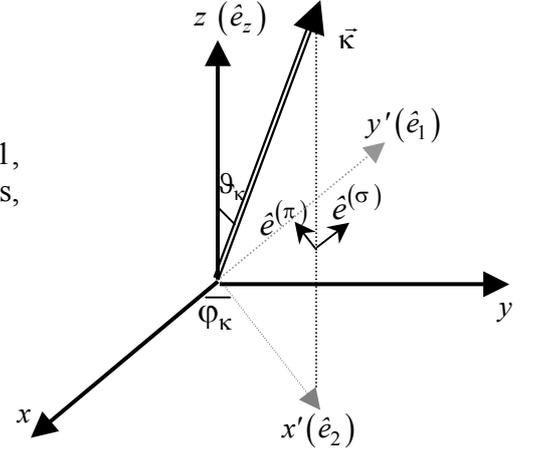

Fig. 1: Coordinate system followed for probabilty amplitude caculation

$$\vec{d} = d_1 \hat{e}_1 + d_2 \hat{e}_2 + d_z \hat{e}_z \qquad (12)$$

with

$$d_1 = \frac{-1}{4V\sqrt{\varepsilon_n \varepsilon_{\bar{n}}}} \sum_{m=-\infty}^{\infty}\sum_{m'=-\infty}^{\infty} c_{m'} c_m^* \int_{-L/2}^{L/2} dz\, e^{i(\kappa_3 - k_3 - k'_3)z} \int_0^\infty \rho d\rho \int_0^{2\pi} d\varphi\, e^{i\kappa_\perp \rho \cos(\varphi-\varphi_\kappa) - i(m+m')\varphi} \times$$

$\{k_\perp[(1-\mathrm{sgn}(\tilde{m}))[e^{i(\varphi_\kappa-\varphi)}Y(|\tilde{m}|-1,|\tilde{m}'|) + e^{-i(\varphi_\kappa-\varphi)}Y(|\tilde{m}|+1,|\tilde{m}'|)] - (1+\mathrm{sgn}(\tilde{m}))[e^{i(\varphi_\kappa-\varphi)}Y(|\tilde{m}|+1,|\tilde{m}'|) +$

$e^{-i(\varphi_\kappa-\varphi)}(|\tilde{m}|-1,|\tilde{m}'|)]] + k'_\perp[(1+\mathrm{sgn}(\tilde{m}'))[e^{-i(\varphi_\kappa-\varphi)}Y(|\tilde{m}|,|\tilde{m}'|-1) + e^{i(\varphi_\kappa-\varphi)}Y(|\tilde{m}|,|\tilde{m}'|+1)] - (1-\mathrm{sgn}(\tilde{m}')) \times$

$[e^{-i(\varphi_\kappa-\varphi)}Y(|\tilde{m}|,|\tilde{m}'|+1) + e^{i(\varphi_\kappa-\varphi)}Y(|\tilde{m}|,|\tilde{m}'|-1)]]\}$,

$$d_2 = \frac{i}{4V\sqrt{\varepsilon_n \varepsilon_{\bar{n}}}} \sum_{m=-\infty}^{\infty}\sum_{m'=-\infty}^{\infty} c_{m'} c_m^* \int_{-L/2}^{L/2} dz\, e^{i(\kappa_3 - k_3 - k'_3)z} \int_0^\infty \rho d\rho \int_0^{2\pi} d\varphi\, e^{i\kappa_\perp \rho \cos(\varphi-\varphi_\kappa) - i(m+m')\varphi} \times$$

$\{k_\perp[(1-\mathrm{sgn}(\tilde{m}))[e^{i(\varphi_\kappa-\varphi)}Y(|\tilde{m}|-1,|\tilde{m}'|) - e^{-i(\varphi_\kappa-\varphi)}Y(|\tilde{m}|+1,|\tilde{m}'|)] + (1+\mathrm{sgn}(\tilde{m}))[e^{-i(\varphi_\kappa-\varphi)}Y(|\tilde{m}|-1,|\tilde{m}'|) -$

$e^{i(\varphi_\kappa-\varphi)}Y(|\tilde{m}|+1,|\tilde{m}'|)]] + k'_\perp[(1+\mathrm{sgn}(\tilde{m}'))[-e^{-i(\varphi_\kappa-\varphi)}Y(|\tilde{m}|,|\tilde{m}'|-1) + e^{i(\varphi_\kappa-\varphi)}Y(|\tilde{m}|,|\tilde{m}'|+1)] - (1-\mathrm{sgn}(\tilde{m}')) \times$

$[e^{i(\varphi_\kappa-\varphi)}Y(|\tilde{m}|,|\tilde{m}'|-1) + e^{-i(\varphi_\kappa-\varphi)}Y(|\tilde{m}|,|\tilde{m}'|+1)]]\}$,

and

$$d_z = \frac{1}{2V\sqrt{\varepsilon_n \varepsilon_{\bar{n}}}} \sum_{m=-\infty}^{\infty}\sum_{m'=-\infty}^{\infty} c_{m'} c_m^* \int_{-L/2}^{L/2} dz\, e^{i(\kappa_3-k_3-k'_3)z} \int_0^\infty \rho d\rho \int_0^{2\pi} d\varphi\, e^{i\kappa_\perp \rho \cos(\varphi-\varphi_\kappa) - i(m+m')\varphi} \times$$

$(k_3 - k'_3) Y(|\tilde{m}|,|\tilde{m}'|).$

In the above expressions, the $Y$-functions are products of two Bessel functions,

$$Y(\alpha,\beta) \equiv J_\alpha(k_\perp \rho) J_\beta(k'_\perp \rho). \qquad (13)$$

Calculation of the probability amplitude integrals is a very involved task. First of all, making use of the box normalization formalism, integration over the *z*-coordinate is worked out from *–L/2* to *L/2* instead of the usual space from $-\infty$ to $\infty$. Secondly, integrations over the φ-coordinate has the following general form:

$$\Xi = \int_0^{2\pi} d(\varphi - \varphi_\kappa) \; e^{-\{i(m+m'\pm 1)\varphi + \kappa\rho \sin\vartheta_\kappa \cos(\varphi-\varphi_\kappa)\}}. \tag{14}$$

It can be calculated by making use of the modified version of Sommerfeld representation of the Bessel function [20], furnishing a third Bessel function. More details are given in Appendix A. The result, with $\theta_\pm \equiv (m+m'\pm 1)$, is

$$\Xi = 2\pi e^{-i\theta_\pm (\varphi_\kappa - \frac{\pi}{2})} J_{-\theta_\pm}(\kappa_\perp \rho). \tag{15}$$

The remaining ρ-integrals are thus reduced to integrals over three Bessel functions of different orders and arguments that show algebraic relationships. They are of two types that can be solved by using tabulated formulae 6.578(3) and 6.522(14) of [21]:

$$\int_0^\infty x J_\nu(b_1 x) J_\mu(b_2 x) J_{\mu+\nu}(cx) dx = 0 \quad \text{for } c > b_1 + b_2 \text{ and } b_1, b_2 > 0. \tag{16-a}$$

$$\int_0^\infty x J_\mu(cx\sin\eta\cos\zeta) J_\nu(cx\cos\eta\sin\zeta) J_{\mu-\nu}(cx) dx = \frac{2}{\pi c^2} \sin(\mu\pi) a^\mu b^\nu D, \tag{16-b}$$

$$\text{for: } c > 0, \; \text{Re } \nu > -1, \; \eta > 0, \; \zeta < \frac{\pi}{2}$$

where $a \equiv \dfrac{\sin\eta}{\cos\zeta}$, $b \equiv \dfrac{\sin\zeta}{\cos\eta}$, $D \equiv [\cos(\eta+\zeta)\cos(\eta-\zeta)]^{-1}$.

In this process, the total energy is conserved as well as the linear momentum along the magnetic tube (z-direction); only the radial momentum is not conserved and satisfies the relation

$$\kappa_\perp > k_\perp + k'_\perp, \tag{17}$$

which means that there is an excess of radial momentum, $\kappa_\perp - (k_\perp + k'_\perp)$, transmitted to the magnetic tube.

Accordingly, we have to set $b_1 = k_\perp$, $b_2 = k'_\perp$, and $c = \kappa_\perp$ in integrals of the type given in Eq.(16-a), and $c\cos\eta \sin\zeta = k_\perp$, $c\sin\eta \cos\zeta = k'_\perp$, $c = \kappa_\perp$ of the type given in Eq.(16-b).

Thus, depending on the indices of Bessel functions, the integral can be directly solved, taking into account the conditions imposed on each integral type as well as the linear dependence properties of Bessel functions, given by $J_{-n}(x) = (-1)^n J_n(x)$, for integer *n*. Then the sums in the probability amplitude components can be evaluated (after we redefine the indices such that $\bar{m} \equiv m + [f]$, $\bar{m}' \equiv m' - [f]$) since they reduce to geometric ones. An example of such a calculation is given in Appendix B.

In carrying out these calculations, an interesting selection rule on the orbital angular momentum quantum numbers $\bar{m}$ and $\bar{m}'$ of the emerging particle-antiparticle pair comes up. An investigation of the conditions after the integrals over the Bessel functions, Eq. (16-a,b), shows that the pair-production process turns out to be forbidden unless the redefined quantum numbers $\bar{m}$ and $\bar{m}'$ of the outgoing particle-antiparticle pair have opposite signs.
Mathematically speaking, for $\bar{m}' \geq 0$, $\bar{m} < 0$ and for $\bar{m}' < 0$, $\bar{m} \geq 0$. Therefore, they have to satisfy the following selection rule:

$$sgn(\bar{m}*\bar{m}') = -1 \tag{18}$$

This means that the created charged particles need to pass the magnetic string in opposite direction. This is necessary for the ingoing photon to transmit the excess of its radial momentum to the string and create the real particle-antiparticle pair from the vacuum. This result was noticed before in the case of spin-1/2 particles [7].

Summing over the redefined indices, the closed form expressions for the probability amplitude takes now the following form:

$$\vec{d} = \frac{D \delta_{k_3, -k_3'} e^{i[f](\varphi_\perp' - \varphi_\perp)} \sin \pi \delta}{L^2 \sqrt{\varepsilon_n \varepsilon_{\bar{n}}} \ \kappa_\perp^2} [i\{A(e^{i\pi\delta}(ab)^\delta \Sigma^+ + e^{-i\pi\delta}(ab)^{-\delta} \Sigma^-)\}\hat{e}_1 +$$

$$\{B(e^{i\pi\delta}(ab)^\delta \Sigma^+ - e^{-i\pi\delta}(ab)^{-\delta} \Sigma^-)\}\hat{e}_2 + \{2(k_3 - k_3')(e^{i\pi\delta}(ab)^\delta \Sigma^+ - e^{-i\pi\delta}(ab)^{-\delta} \Sigma^-)\}\hat{e}_z].$$

(19)

where

$$\Sigma^+ \equiv \frac{1}{1 - ae^{-i\varphi_{\perp\kappa}}} \frac{be^{-i\varphi'_{\perp\kappa}}}{1 - be^{-i\varphi'_{\perp\kappa}}}, \qquad \Sigma^- \equiv \frac{ae^{i\varphi_{\kappa\perp}}}{1 - ae^{i\varphi_{\kappa\perp}}} \frac{1}{1 - be^{i\varphi'_{\perp\kappa})}}, \qquad (20)$$

$$A \equiv k_\perp(a - a^{-1}) + k_\perp'(b - b^{-1}), \qquad B \equiv k_\perp(a - a^{-1}) - k_\perp'(b - b^{-1}), \qquad (21)$$

$$a(k_\perp, k_\perp', \kappa_\perp) = \frac{2k_\perp \kappa_\perp}{(k_\perp'^2 + \kappa_\perp^2 - k_\perp^2) + \sqrt{k_\perp^4 - 2\kappa_\perp^2(k_\perp'^2 + k_\perp^2) + (k_\perp'^2 - k_\perp^2)^2}}, \qquad \text{(22-a)}$$

$$b(k_\perp, k_\perp', \kappa_\perp) = \frac{2k_\perp' \kappa_\perp}{(k_\perp'^2 + \kappa_\perp^2 - k_\perp^2) + \sqrt{k_\perp^4 - 2\kappa_\perp^2(k_\perp'^2 + k_\perp^2) + (k_\perp'^2 - k_\perp^2)^2}}, \qquad \text{(22-b)}$$

$$D = \frac{\kappa_\perp^2 \ ab}{k_\perp k_\perp'[1 - a^2 b^2]}, \quad \text{and} \qquad (23)$$

$$\varphi_{ij} \equiv \varphi_i - \varphi_j. \qquad (24)$$

In Eq.(19), we considered only the case of normal incidence of the incoming photon on the solenoid, i.e. $\vartheta_\kappa = \pi/2$. This simplifies subsequent calculations without introducing any loss of generality.

For pair production process by a single photon of specific energy, the *transition probability per unit time* is given by [19]:

$$w_{\bar{n}n} = \frac{4\pi^2 e^2}{\hbar \kappa L^3} \sum_{n_3, n_3'} (\vec{d}^* \cdot \hat{a}^+)(\vec{d} \cdot \hat{a}) \delta(\varepsilon_{\bar{n}} + \varepsilon_n - \kappa) \qquad (25)$$

where the sum is evaluated over the z-components of the momenta of the particles and antiparticles, respectively. Summation over $n_3$ and $n_3'$ can be performed with the help of the formula [19]

$$\sum_{n_3, n_3'} \delta_{k_3, -k_3'} = \sum_{n_3, n_3'} \delta_{n_3, -n_3'} = \frac{L}{2\pi} \int dk_3. \qquad (26)$$

It is clear that the z-components of the momenta of the created emitted pair are equal in magnitude, but have opposite directions.

If the amplitudes are expanded in terms of the linear polarization vectors, $\hat{\beta}_\lambda$, described in Fig.1 such that

$$\hat{\beta}_\lambda = \begin{cases} \hat{e}^{(\sigma)} = (-\sin\varphi_\kappa,\ \cos\varphi_\kappa,\ 0) & \text{for } \lambda = 2 \\ \hat{e}^{(\pi)} = (-\cos\vartheta_\kappa\cos\varphi_\kappa,\ -\cos\vartheta_\kappa\sin\varphi_\kappa,\ \sin\vartheta_\kappa) & \text{for } \lambda = 3 \end{cases} \quad (27)$$

then the transition probability per unit time for a given polarization state $\lambda$ reads

$$w_{\bar{n}n,\lambda} = \frac{2\pi e^2}{\hbar\kappa}\frac{1}{L^2}\int_{-\infty}^{\infty} dk'_3\ \Lambda_\lambda \delta(\varepsilon_{\bar{n}} + \varepsilon_n - \kappa) \quad (28)$$

with

$$\Lambda_\lambda = (\vec{d}^* \cdot \hat{\beta}_\lambda)(\vec{d}\cdot\hat{\beta}_\lambda). \quad (29)$$

The calculations will be done with respect to the rotated coordinate system, $(x', y')$ (see Fig. 1), in which $\varphi_\kappa = 0$, then $\hat{e}^{(\sigma)} = (0,\ 1,\ 0)$ and $\hat{e}^{(\pi)} = (-\cos\vartheta_\kappa, 0, \sin\vartheta_\kappa)$. In other words, the polarization vector may be taken to be *normal* to the plane determined by $\vec{\kappa}$ and $\hat{e}_2$ for $\hat{e}^{(\sigma)}$, while for $\hat{e}^{(\pi)}$ it must then lie *in* that plane. For normal incidence, it is not difficult to show that

$$\Lambda^{(\sigma)} = \frac{D^2 \sin^2\pi\delta}{L^4 \varepsilon_n \varepsilon_{\bar{n}} \kappa^4} A^2 \{(ab)^{2\delta}|\Sigma^+|^2 + (ab)^{-2\delta}|\Sigma^-|^2 + e^{2i\pi\delta}\Sigma^+\Sigma^{-*} + e^{-2i\pi\delta}\Sigma^-\Sigma^{+*}\}. \quad (30\text{-a})$$

$$\Lambda^{(\pi)} = \frac{D^2 \sin^2\pi\delta}{L^4 \varepsilon_n \varepsilon_{\bar{n}} \kappa^4} 4(k_3 - k'_3)^2 \{(ab)^{2\delta}|\Sigma^+|^2 + (ab)^{-2\delta}|\Sigma^-|^2 - e^{2i\pi\delta}\Sigma^+\Sigma^{-*} - e^{-2i\pi\delta}\Sigma^-\Sigma^{+*}\}, \quad (30\text{-b})$$

where

$$|\Sigma^+|^2 = \frac{b^2}{[1-2a\cos\varphi_{\kappa\perp}+a^2][1-2b\cos\varphi'_{\perp\kappa}+b^2]},\quad |\Sigma^-|^2 = \frac{a^2}{[1-2a\cos\varphi_{\kappa\perp}+a^2][1-2b\cos\varphi'_{\perp\kappa}+b^2]}.$$

Finally, the complete information about energy, angular, and polarization distributions of created particles and antiparticles is contained in the effective differential cross section given by

$$d\sigma_\lambda = \Gamma(\varepsilon_n, \varepsilon_{\bar{n}})\frac{1}{J}dw_{n\bar{n}}. \quad (31)$$

where

$$\Gamma(\varepsilon_n, \varepsilon_{\bar{n}}) = \left(\frac{1}{c}\right)\left(\frac{L}{2\pi}\right)^4 k_\perp dk_\perp d\varphi_\perp k'_\perp dk'_\perp d\varphi'_\perp, \quad (32)$$

and

$$J = \frac{c}{L^2}. \quad (33)$$

Consequently,

$$\frac{d\sigma_\lambda}{dk_\perp d\varphi_\perp dk'_\perp d\varphi'_\perp dk'_3} = \frac{2\pi e^2}{\hbar\kappa}\left(\frac{L}{2\pi}\right)^4 k_\perp k'_\perp \Lambda_\lambda\big|_{\varepsilon_n + \varepsilon_{\bar{n}} = \kappa}. \quad (34)$$

where $\Lambda_\lambda$ are given in Eqs(30-a,b), and $c$ is the speed of light in vacuum.

## IV. Limiting Cases:
The above differential cross-section is somewhat of complicated form. So it is worthwhile to illustrate limiting cases.

A. Non-relativistic Limit

In this limit, the incident photon energy is just above the threshold energy. i.e. $\hbar\omega - 2Mc^2 \simeq 0$. So, $\kappa \sim 2Mc/\hbar$ and $k_\perp \sim k'_\perp \sim k_3 \sim k'_3 \ll Mc/\hbar$. Then the nonrelativistic (NR) version of Eqs. (21-23) reduces to

$$a_{NR} \simeq \frac{k_\perp}{\kappa} = \frac{\hbar k_\perp}{2Mc}, \quad b_{NR} \simeq \frac{k'_\perp}{\kappa} = \frac{\hbar k'_\perp}{2Mc}, \quad D_{NR} \simeq 1, \quad A_{NR} \simeq -2\kappa = \frac{-4Mc}{\hbar}, \quad B_{NR} \simeq 0,$$

By substituting the above approximate values, the probability amplitude is considerably simplified. Its components read

$$d_1^{NR} \simeq \frac{-i\hbar^3 e^{i[f]\varphi_{\perp'\perp}}\sin(\pi\delta)}{2L^2 M^3 c^3}\left\{\left(\frac{\hbar^2 k_\perp k'_\perp}{4M^2 c^2}\right)^\delta k'_\perp e^{-i\varphi_{\perp'\kappa}+i\pi\delta} + \left(\frac{\hbar^2 k_\perp k'_\perp}{4M^2 c^2}\right)^{-\delta} k_\perp e^{-i\varphi_{\perp\kappa}-i\pi\delta}\right\}, \quad (35\text{-}a)$$

$$d_2^{NR} \simeq 0, \quad (35\text{-}b)$$

$$d_z^{NR} \simeq \frac{\hbar^4 k_3 e^{i[f]\varphi_{\perp'\perp}}\sin(\pi\delta)}{2L^2 M^4 c^4}\left\{\left(\frac{\hbar^2 k_\perp k'_\perp}{4M^2 c^2}\right)^\delta k'_\perp e^{-i\varphi_{\perp'\kappa}+i\pi\delta} + \left(\frac{\hbar^2 k_\perp k'_\perp}{4M^2 c^2}\right)^{-\delta} k_\perp e^{-i\varphi_{\perp\kappa}-i\pi\delta}\right\}. \quad (35\text{-}c)$$

And

$$\Lambda_{NR}^{(\sigma)} \simeq \frac{\hbar^6 \sin^2(\pi\delta)}{4L^4 M^6 c^6}\left\{\left(\frac{\hbar^2 k_\perp k'_\perp}{4M^2 c^2}\right)^{2\delta} k'^2_\perp + \left(\frac{\hbar^2 k_\perp k'_\perp}{4M^2 c^2}\right)^{-2\delta} k^2_\perp\right\},$$

$$\Lambda_{NR}^{(\pi)} \simeq \frac{\hbar^8 k_3^2 \sin^2(\pi\delta)}{4L^4 M^8 c^8}\left\{\left(\frac{\hbar^2 k_\perp k'_\perp}{4M^2 c^2}\right)^{2\delta} k'^2_\perp + \left(\frac{\hbar^2 k_\perp k'_\perp}{4M^2 c^2}\right)^{-2\delta} k^2_\perp\right\}.$$

Eventually,

$$\frac{d\sigma_{NR}^{(\sigma)}}{dk_\perp d\varphi_\perp dk'_\perp d\varphi'_\perp dk'_3} \simeq \frac{e^2 \hbar^5 k_\perp k'_\perp \sin^2(\pi\delta)}{(2\pi)^3 4\kappa M^6 c^6}\left\{\left(\frac{\hbar^2 k_\perp k'_\perp}{4M^2 c^2}\right)^{2\delta} k'^2_\perp + \left(\frac{\hbar^2 k_\perp k'_\perp}{4M^2 c^2}\right)^{-2\delta} k^2_\perp\right\}, \quad (36\text{-}a)$$

$$\frac{d\sigma_{NR}^{(\pi)}}{dk_\perp d\varphi_\perp dk'_\perp d\varphi'_\perp dk'_3} \simeq \frac{e^2 \hbar^7 k_\perp k'_\perp k_3^2 \sin^2(\pi\delta)}{(2\pi)^3 4\kappa M^8 c^8}\left\{\left(\frac{\hbar^2 k_\perp k'_\perp}{4M^2 c^2}\right)^{2\delta} k'^2_\perp + \left(\frac{\hbar^2 k_\perp k'_\perp}{4M^2 c^2}\right)^{-2\delta} k^2_\perp\right\}. \quad (36\text{-}b)$$

A remarkable feature of the cross section in the nonrelativistic limit can be read off the above two equations: Unlike the spinor case [7], the differential scattering cross section for the $\sigma$-polarization is much larger than the $\pi$-polarization. It is larger by an order that is proportional to $(Mc/\hbar k_3)^2 \gg 1$. Physically speaking, it means that it is unlikely that the low-energy $\pi$-polarized photon creates a scalar particle-antiparticle pair but it is created mainly by the $\sigma$-polarized photon.

B. Ultrarelativistic Limit:

If the photon energy is much larger than the threshold energy, i.e. $\hbar c\kappa \gg 2Mc^2$, then the created pair will be emitted predominantly in the forward direction within too narrow cone about the direction of the incident photon. This means that the third components of the linear momentum, $k_3$ and $k'_3$, of the outgoing pair are very small and can be neglected. In the lowest order approximation, the pair energy is approximately of kinetic type, i.e. $\varepsilon_n \approx \sqrt{k_3^2 + k_\perp^2} \approx k_\perp$ and $\varepsilon_{\bar{n}} \approx \sqrt{k_3'^2 + k_\perp'^2} \approx k_\perp'$.

Accordingly, it is not difficult to conclude that $\varphi_\kappa \cong \varphi_\perp \cong \varphi_{\perp'}$. Then the angular distribution of the emitted particle-antiparticle pair, in the ultrarelativistic limit (UR) is simplified considerably. In this limit, we have $a_{UR} \simeq \frac{2k}{\kappa}, b_{UR} \simeq \frac{2k'}{\kappa}, A_{UR} \simeq -\kappa, D_{UR} \simeq 4$ and

$$\Lambda_{UR}^{(\sigma)} \simeq \frac{16 \sin^2 \pi \delta}{L^4 k k' \kappa^2} (\Sigma^+)^2 \left\{ \left( \frac{4 k k'}{\kappa^2} \right)^{2\delta} + \frac{k^2}{k'^2} \left( \frac{4 k k'}{\kappa^2} \right)^{-2\delta} + \frac{2k}{k'} \cos 2\pi\delta \right\}, \tag{37-a}$$

$$\Lambda_{UR}^{(\pi)} \simeq \frac{(16 k_3)^2 \sin^2 \pi \delta}{L^4 k k' \kappa^4} (\Sigma_{UR}^+)^2 \left\{ \left( \frac{4 k k'}{\kappa^2} \right)^{2\delta} + \frac{k^2}{k'^2} \left( \frac{4 k k'}{\kappa^2} \right)^{-2\delta} - \frac{2k}{k'} \cos 2\pi\delta \right\}. \tag{37-b}$$

where

$$\Sigma_{UR}^+ = \Sigma_{UR}^{+*} \simeq \frac{b_{UR}}{(1-b_{UR})(1-a_{UR})}, \tag{38-a}$$

$$\Sigma_{UR}^- = \Sigma_{UR}^{-*} = \frac{a_{UR}}{b_{UR}} \Sigma_{UR}^+ \simeq \frac{a_{UR}}{(1-b_{UR})(1-a_{UR})}. \tag{38-b}$$

In contrast to the nonrelativistic limit, it can be easily recognized from Eqs. (37-a,b), that in the ultrarelativistic limit, the differential cross section for $\pi$-polarization is much smaller than the $\sigma$-polarization. It is smaller by an order that is proprtional to $(k_3/\kappa)^2$.

*V. Conclusions*
       We have analysed the scalar pair production by a single, high-energy linearly polarized photon in the presence of *AB* potential. In this case, photons do not intertact directly with the magnetic field because they cannot penetrate the magnetic string, and the process hasppens due to the purely non-local interaction of the created charged particles with the *AB* vector potential, in contrast to the spin-1/2 case, where the interaction has a local part.

       The first order differetial scattering cross section was calculated within the framework of time-dependent perturbation theory using, however, the exact solutions of the Klein-Gordon equation as expansion basis of the field operators.. We showed that the process turns out to be forbidden unless the quantum numbers $\bar{m}$ and $\bar{m}'$ of the outgoing particle and antiparticle have opposite signs. This means that the created virtual pair encircle the magnetic string in opposite directions, in order to be transformed into a real one. In the non-relativistic limit, the cross section for the scalar particle-antiparticle pair production by $\sigma$-polarized photons was found to be much greater than that by $\pi$-polarized photons. This is in contrast to the case of spinor particle-antiparticle pair production, where the dominant cross section is that of $\pi$-polarized photons [7].

# Appendix A

The $\varphi$-integral, given in Eq. (14) has the following general form:

$$\Xi = \int_0^{2\pi} d(\varphi - \varphi_\kappa) \; e^{-i\{(m+m'\pm 1)\varphi - \kappa_\perp \rho \cos(\varphi - \varphi_\kappa)\}} . \tag{A-1}$$

Defining new variables, $\varphi - \varphi_\kappa = \chi + \pi$, $\theta_\pm \equiv m + m' \pm 1$, and $\bar{\varphi}_\kappa \equiv \varphi_\kappa - \dfrac{\pi}{2}$, the inegral reduces to

$$\Xi = e^{-i\theta_\pm(\varphi_\kappa - \pi)} \int_{-\pi}^{\pi} d\chi \; e^{i\theta_\pm \chi + i\kappa_\perp \rho \cos\chi} . \tag{A-2}$$

Next, one can make use of the Sommerfeld's representation of Bessel function of order $\nu$ given by

$$J_\nu(z) = \frac{e^{i\pi\nu/2}}{2\pi} \int_C dt\, e^{-iz\cos t + i\nu t} , \tag{A-3}$$

where $C$ is the contour that goes from $-\pi + \xi + i\infty$ to $\pi + \xi + i\infty$ with $\xi$ is a positive infinitesimal, as illustrated in Fig. 2.

Eventually, the $\varphi$-integral can be easily simplified in terms of Bessel funcion, with the result

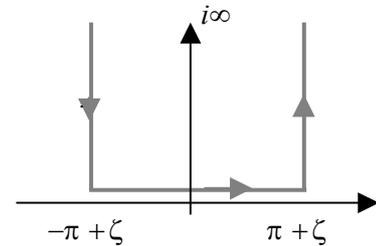

Fig. 2: Sommerfeld's representaion of Bessel function.

$$\Xi = 2\pi\, e^{-i\theta_\pm (\varphi_\kappa - \frac{\pi}{2})} J_{-\theta_\pm}(\kappa_\perp \rho) . \tag{A-4}$$

# Appendix B

In this Appendix, a full sample calculation of the first $\rho$-integral in $d_1$ component of Eq. (12) will be worked out in details.

Let

$$G_1 = \sum_{m=-\infty}^{\infty} \sum_{m'=-\infty}^{\infty} c_{m'} c_m^* \, e^{-i(m+m')\bar{\varphi}_\kappa} \left[1 - \text{sgn}(\tilde{m})\right] \int_0^\infty \rho \, d\rho \int_0^{2\pi} d\varphi \, e^{i\kappa_\perp \rho \cos(\varphi - \varphi_\kappa) - i(m+m')\varphi} \times$$

$$(1 - \text{sgn}(\tilde{m})) e^{i(\varphi_\kappa - \varphi)} Y(|\tilde{m}| - 1, |\tilde{m}'|).$$

Using the result of Eq. (A-4) for the $\varphi$-integral and Eq. (13) with $\alpha = |\tilde{m}| - 1, \beta = |\tilde{m}'|$, then

$$G_1 = \sum_{m=-\infty}^{\infty} \sum_{m'=-\infty}^{\infty} c_{m'} c_m^* \, e^{-i(m+m')\bar{\varphi}_\kappa} \left[1 - \text{sgn}(\tilde{m})\right] \int_0^\infty \rho \, d\rho \, J_{|\tilde{m}|-1}(k_\perp \rho) J_{|\tilde{m}'|}(k'_\perp \rho) J_{\theta_+}(\kappa_\perp \rho) \qquad (B.1)$$

Then the summation over indices $m$ and $m'$ should be redefined as follows:

$$\left. \begin{array}{l} \bar{m} = m + [f] \Rightarrow \tilde{m} = \bar{m} + \delta \\ \bar{m}' = m' - [f] \Rightarrow \tilde{m}' = \bar{m}' - \delta \end{array} \right\} \Rightarrow m + m' = \bar{m} + \bar{m}'. \qquad (B.2)$$

Next, summing over the new indices and partitioning both summations for positive and negative indices values, it appears that $G_1$ can be partitioned into four terms. These terms will be designated by the letters $T_1, T_2, T_3$, and $T_4$, respectively.

$$G_1 = \sum_{\bar{m}=-\infty}^{-1} \sum_{\bar{m}'=-\infty}^{-1} \mathfrak{A}(\bar{m}, \bar{m}') + \sum_{\bar{m}=-\infty}^{-1} \sum_{\bar{m}'=0}^{\infty} \mathfrak{A}(\bar{m}, \bar{m}') + \sum_{\bar{m}=0}^{\infty} \sum_{\bar{m}'=-\infty}^{-1} \mathfrak{A}(\bar{m}, \bar{m}') + \sum_{\bar{m}=0}^{\infty} \sum_{\bar{m}'=0}^{\infty} \mathfrak{A}(\bar{m}, \bar{m}'), \qquad (B3)$$

where

$$\mathfrak{A}(\bar{m}, \bar{m}') \equiv c_{\bar{m}}^* c_{\bar{m}'} e^{-i(\bar{m}+\bar{m}')\bar{\varphi}_\kappa} \left[1 - \text{sgn}(\bar{m} + \delta)\right] \int_0^\infty \rho \, d\rho \, J_{|\tilde{m}|-1}(k_\perp \rho) J_{|\tilde{m}'|}(k'_\perp \rho) J_{\theta_+}(\kappa_\perp \rho). \qquad (B.4)$$

First Term: $T_1$

$$T_1 = \sum_{\bar{m}=-\infty}^{-1} \sum_{\bar{m}'=-\infty}^{-1} c_{\bar{m}}^* c_{\bar{m}'} e^{-i(\bar{m}+\bar{m}')\bar{\varphi}_\kappa} \left[1 - \text{sgn}(\bar{m} + \delta)\right] \int_0^\infty \rho \, d\rho \, J_{|\bar{m}+\delta|-1}(k_\perp \rho) J_{|\bar{m}'-\delta|}(k'_\perp \rho) \, J_{\bar{m}+\bar{m}'+1}(\kappa_\perp \rho).$$

For $\bar{m} < 0 \; \& \; \bar{m}' < 0 : |\bar{m} + \delta| = -\bar{m} - \delta \; , |\bar{m}' - \delta| = -\bar{m}' + \delta$.

Making use of the fact that $J_{-n}(x)$ and $J_n(x)$ are "linearly dependent" for integer index $n$, then

$$T_1 = -2 \, e^{i[f](\varphi_{\perp'} - \varphi_\perp)} \sum_{\bar{m}=-\infty}^{-1} e^{i\bar{m}(\varphi_\perp - \varphi_\kappa - \pi/2)} (-1)^{-\bar{m}} \sum_{\bar{m}'=-\infty}^{-1} e^{i\bar{m}'(\varphi'_\perp - \varphi_\kappa - \pi/2)} (-1)^{-\bar{m}'} \times$$

$$\int_0^\infty d\rho \, \rho \, J_{-\bar{m}-\delta-1}(k_\perp \rho) J_{-\bar{m}'+\delta}(k'_\perp \rho) J_{-\bar{m}-\bar{m}'-1}(\kappa_\perp \rho).$$

An investigation of the integral in $T_1$ expression shows that it must vanish since it is an integral of Eq.(16-a) type, for $\mu = -\bar{m} - \delta - 1$ and $\nu = -\bar{m}' + \delta$. Therefore, with $\kappa_\perp > k_\perp + k'_\perp$,

$$(-\bar{m} - \delta - 1) + (-\bar{m}' + \delta) = (-\bar{m} - \bar{m}' - 1) = \mu + \nu$$

Second Term: $T_2$

$$T_2 = \sum_{\bar{m}=-\infty}^{-1} \sum_{\bar{m}'=0}^{\infty} c_{\bar{m}}^* c_{\bar{m}'} e^{-i(\bar{m}+\bar{m}')\bar{\varphi}_\kappa} [1-\text{sgn}(\bar{m}+\delta)] \int_0^\infty \rho \, d\rho \, J_{|\bar{m}+\delta|-1}(k_\perp \rho) J_{|\bar{m}'-\delta|}(k'_\perp \rho) \, J_{\bar{m}+\bar{m}'+1}(\kappa_\perp \rho).$$

It is not difficult to verify that for $\bar{m}<0$ & $\bar{m}'>0$, the above integral is of type Eq.(16-b) with $\nu = -\bar{m} - \delta - 1$ and $\mu = \bar{m}' - \delta$ for $\kappa_\perp > k_\perp + k'_\perp$. Thus

$$T_2 = \frac{4D}{\pi \, \kappa_\perp^2} a^{-\delta-1} b^{-\delta} e^{i[f](\varphi'_\perp - \varphi_\perp)} e^{-i\pi\delta} \sum_{\bar{m}=-\infty}^{-1} a^{-\bar{m}} e^{i\bar{m}(\varphi_\perp - \varphi_\kappa - \pi)} \sin[\pi(m+\delta+1)] \sum_{\bar{m}'=0}^{\infty} b^{\bar{m}'} e^{i\bar{m}'(\varphi'_\perp - \varphi_\kappa)}.$$

Notice that the two summations are turned out to be geometric.

$$\sum_{\bar{m}'=0}^{\infty} b^{\bar{m}'} e^{i\bar{m}'(\varphi'_\perp - \varphi_\kappa)} = \frac{1}{1-be^{i(\varphi'_\perp - \varphi_\kappa)}}.$$

Likewise,

$$\sum_{\bar{m}=-\infty}^{-1} a^{-\bar{m}} e^{i\bar{m}(\varphi_\perp - \varphi_\kappa - \pi)} \sin[\pi(m+\delta+1)] = \sum_{\bar{m}=-\infty}^{-1} a^{-\bar{m}} e^{i\bar{m}(\varphi_\perp - \varphi_\kappa)} \left( \frac{e^{i\pi\delta} - e^{-i\pi\delta}}{2i} \right)$$

$$= \sum_{\bar{\ell}=1}^{\infty} a^{\bar{\ell}} e^{-i\bar{\ell}(\varphi_\perp - \varphi_\kappa)} \sin\pi\delta = \frac{\sin\pi\delta \, ae^{-i(\varphi_\perp - \varphi_\kappa)}}{1-ae^{-i(\varphi_\perp - \varphi_\kappa)}},$$

where we have set $\bar{\ell} \equiv -\bar{m}$. Finally,

$$T_2 = \frac{4D\sin\pi\delta}{\pi \kappa_\perp^2} e^{i[f](\varphi'_\perp - \varphi_\perp)} (ab)^{-\delta} e^{-i\pi\delta} \frac{e^{-i(\varphi_\perp - \varphi_\kappa)}}{1-ae^{-i(\varphi_\perp - \varphi_\kappa)}} \frac{1}{1-be^{i(\varphi'_\perp - \varphi_\kappa)}}. \quad \text{(B.5)}$$

Third Term: $T_3$

$$T_3 = \sum_{\bar{m}=0}^{\infty} \sum_{\bar{m}'=-\infty}^{-1} c_{\bar{m}}^* c_{\bar{m}'} \, e^{-i(\bar{m}+\bar{m}')\bar{\varphi}_\kappa} [1-\text{sgn}(\bar{m}+\delta)] \int_0^\infty \rho \, d\rho \, J_{|\bar{m}+\delta|-1}(k_\perp \rho) J_{|\bar{m}'-\delta|}(k'_\perp \rho) J_{\bar{m}+\bar{m}'+1}(\kappa_\perp \rho).$$

For $\bar{m}>0$ & $\bar{m}'<0$: $|\bar{m}+\delta|=\bar{m}+\delta$, $|\bar{m}'-\delta|=-\bar{m}'+\delta$. Also, we have $\text{sgn}(\bar{m}+\delta)=1$. Therefore, the third term $T_3$ vanishes.

Fourth Term: $T_4$

$$T_4 = \sum_{\bar{m}=0}^{\infty} \sum_{\bar{m}'=0}^{\infty} c_{\bar{m}}^* c_{\bar{m}'} e^{-i(\bar{m}+\bar{m}')\bar{\varphi}_\kappa} [1-\text{sgn}(\bar{m}+\delta)] \int_0^\infty \rho \, d\rho \, J_{|\bar{m}+\delta|-1}(k_\perp \rho) J_{|\bar{m}'-\delta|}(k'_\perp \rho) J_{\bar{m}+\bar{m}'+1}(\kappa_\perp \rho).$$

For $\bar{m}>0$ & $\bar{m}'>0$: $|\bar{m}+\delta|-1=\bar{m}+\delta-1$, $|\bar{m}'-\delta|=\bar{m}'-\delta$. Also, we have $[\text{sgn}(\bar{m}+\delta)-1]=0$.

Consequently, the fourth term $T_4$ must vanish as well, and that only one term survives in the $G_1$ expression above.